\documentstyle[12pt]{article}

\begin{document}

\newcommand{\sect}{\section}

\renewcommand {\theequation}{\thesection.\arabic{equation}}
\renewcommand{\sect}[1]{\section{#1}\setcounter{equation}{0} }

\newcommand{\eqr}{\begin{eqnarray}}
\newcommand{\rqe}{\end{eqnarray}}

\newcommand{\eq}{\begin{equation}} 
\newcommand{\qe}{\end{equation}}




\begin{flushright}
HUTP-98/A022\\
hep-th/9805203
\end{flushright}

\vspace{2cm}

\begin{center}

{\bf\Large
(Supersymmetric) Kac-Moody Gauge Fields in 3+1 Dimensions
} 

\vspace{1.5cm}

Belal E. Baaquie\footnote{Permanent address: Department of Physics, 
National University of Singapore, Kent Ridge Road, Singapore 
091174; e-mail:phybeb@nus.edu.sg} 

\vspace{0.2cm}

{\em
Lyman Laboratory of Physics\\
Harvard University\\
Cambridge, MA 02138, U.S.A.
}
\end{center}

\vspace{1.4cm}

\begin{abstract}

Lagrangians for gauge fields and matter fields can be constructed from 
the infinite dimensional Kac-Moody algebra and group.   A continuum
regularization is used to obtain such generic  
lagrangians, which contain new nonlinear and asymmetric
interactions not present in gauge theories based on compact Lie
groups. This technique is applied to deriving the Yang-Mills and
Chern-Simons lagrangians for the Kac-Moody case.  The extension of this
method to $D=4, N=({1\over2},0)$ supersymmetric Kac-Moody gauge fields is
also made. 

\end{abstract}

\thispagestyle{empty}

\newpage

\setcounter{page}{1}

\sect{Introduction}

Gauge fields and matter fields having a local gauge invariance given
by the Kac-Moody group have been written down in \cite{b1pl}.  It was
shown in \cite{b2pr} that $\hat{U}(1)$  Kac-Moody fermions exhibit a maximum
limiting temperature and it was further shown in \cite{r} that in one
loop the  $\hat{U}(1)$ Kac-Moody gauge field in D=3+1 is renormalizable as well
as asymptotically free.

The lagrangian for Kac-Moody gauge fields was motivated in \cite{b1pl}
using ideas from lattice gauge theory. The result was stated for the
bosonic sector without any derivation, and the regularization required for the
Kac-Moody gauge group was only addressed for the case of Kac-Moody
fermions. In this paper we show how to derive the result directly from
the continuum formulation and it will then become clear how the
lagrangian for Kac-Moody gauge fields depends crucially on using a
regularization.

A general feature of Kac-Moody gauge fields is that the central extension
yields two sets of spacetime gauge fields, namely one set being in general a
non-Abelian bulk gauge field $A_\mu$ defined on a $D=d+1$-dimensional
manifold and a $U(1)$ boundary gauge field $B_i$ defined on a
$d$-dimensional manifold. The central extension couples these two fields in
a gauge and supersymmetric invariant manner.

In Section 2 Kac-Moody gauge fields and field tensor, gauge
transformations, the regularization as well as the generalization of
the Wilson line to a Wilson cylinder are discussed.  In Section 3 the
Yang-Mills lagrangian for the Kac-Moody case is derived and in Section
4 the lagrangian for the Chern-Simons theory is obtained.  In Section 5
the results are generalized to the case of supersymmetric Kac-Moody
gauge fields and in Section 6 some conclusions are drawn.

\sect{Kac-Moody Gauge Fields}

Let $Q_a$ be the generators of the Kac-Moody algebra based on underlying
compact Lie group $G$ satisfying the
commutation equations ($' = \partial/{\partial\sigma}$)

\eq
\label{ccr}
[Q_a(\sigma),Q_b(\sigma^\prime)]=iC_{abc}\delta(\sigma-\sigma^\prime)Q_c(\sigma)+
ik\delta^{ab}\delta'(\sigma - \sigma^\prime) 
\qe

where $\sigma,\sigma^\prime \in S^1$, and $k$=integer/2$\pi$. For $k=0$ we
recover the loop group algebra. 

Gauge transformations $\Phi(x)$ are given by finite elements of the Kac-Moody group $\hat{G}$.  For $\int \equiv \int_{0}^{R} d\sigma$ (that is
 ${\cal S}^{1}$ has radius $ R /2\pi$) we have

\eq
\label{gtp}
\Phi(x)=exp\{i\Lambda(x)+i\int\phi^{a}(x,\sigma)Q_{a}(\sigma)\}
\qe

with $x\in M^3$ being a element of 3-dimensional Minkowski spacetime.

Define the Kac-Moody gauge field by

\eq
{\cal A}_i(x)=eB_i(x) + g\int A^a_i(x,\sigma)Q_a(\sigma)
\qe

Note $B_i(x)$ is a $U(1)$ gauge field due to the central extension defined
on $M^3$
and $A^a_i(x,\sigma)$ is a non-Abelian gauge field defined on spacetime
$M^3 \times S^1$. We will refer to the fields on $M^3$ as boundary  fields
and fields on $M^3 \times S^1$ as bulk fields.  On manifold $M^d\times S^1$ the coupling constant g has mass dimension (d-2)/2 and e has mass dimension of (d-3)/2. For simplicity, from now on we will set all couplings to
unity. The indices $i,j,k$ etc run through 1,2,3 and indices $\mu,\nu$ etc run
through 1,2,3,4 where $x_4=\sigma$. 

Gauge transformations are given by

\eq
\label{gt}
{\cal A}^{\Phi}_i(x)= \Phi(x)
{\cal A}_i(x)\Phi^\dag (x)+i\Phi(x)\partial_i\Phi^\dag(x)
\qe

To obtain the gauge transformation in terms of $B_i$ and
$A^a_i$  note that \cite{b1pl} we have

\eq
\label{gtQ}
\Phi(x)Q_a(\sigma)\Phi^\dag(x)=\rho^T_{ab}(x,\sigma)Q_b(\sigma)+ke^T_{ab}(x,\sigma)\phi'_b(x,\sigma)
\qe

Let $T^a$ be the generators of  $G$; define matrix
$\phi(x,\sigma)=exp\{\phi_a(x,\sigma)T^a\}\in G$;  the vierbien on G is
given by 
$e_{ab}(x,\sigma)T^b=-i\phi^{\dag}\partial_a\phi$ and the adjoint matrix
by $\phi^{\dag}T_a \phi=\rho_{ab} T^b$.
 
Using the functional differential realization of the Kac-Moody generators
\cite{bkm} we have 

\eqr 
\label{diff}
-\imath\partial_i
\Phi(x)\Phi^\dag(x)&=&\int\rho_{ab}(x,\sigma)\partial_i\phi_a(x,\sigma)Q_b(\sigma)
\nonumber\\
  &-&k\int G_{ab}(x,\sigma)\partial_i\phi_a(x,\sigma)\phi'_b(x,\sigma)
\rqe

where $G_{ab}$ is given in \cite{b1pl}.

Hence from the equations given above we obtain for $A_i=A^a_iT^a$,

\eq
\label{gta}
 A_i^\phi(x,\sigma)= \phi A_i\phi^\dag +i\phi\partial_i\phi^\dag(x,\sigma)
\qe

which is the usual gauge transformation for compact Lie groups; the non-trivial 
transformation due to the central extension is given by  \cite{b1pl}  

\eq
\label{gtb}
B_i^\Omega= B_i - 
\partial_i\Lambda +k\int e_{ab}\phi'_bA^a_i -k\int G_{ab}\partial_i\phi_a\phi_b^\prime 
\qe

Define the Yang-Mills field tensor for the Kac-Moody case by

\eq
\label{ym}
{\cal F}_{ij}=\partial_i{\cal A}_j - \partial_j{\cal A}_i + i[{\cal
A}_i,{\cal A}_j]
\qe

and which transforms covariantly under Kac-Moody gauge transformations.
Introduce a new dynamical 
nonlinear scalar field $\Omega \in \hat{G}$  defined by

\eq
\label{omega}
\Omega(x)= exp\{i\alpha(x)+i\int\omega^a(x,\sigma)Q_a(\sigma)\}
\qe   

Define

\eq
\label{ymomega}
{\cal F}^\Omega_{ij}(x)=\Omega(x){\cal F}_{ij}(x)\Omega^\dag(x)
\qe
In components for, $f_{ij}=\partial_iB_j-\partial_jB_i$, we have

\eq
\label{Fij1}
{\cal F}_{ij}(x)=f_{ij}+k\int A'^a_i A^a_j +k\int F^a_{ij}(x,\sigma)Q_a(\sigma)
\qe

with $F_{ij}= \partial_i A_j-\partial_j A_i +i[A_i,A_j]$.

We also have, from (\ref{ymomega}) and (\ref{Fij1}) that 

\eq
\label{Fij2}
{\cal F}_{ij}^\Omega= \Gamma_{ij}(x)+\int F^a_{ij}(x,\sigma)Q_b(\sigma)\rho^T_{ab}(\sigma)
\qe

where

\eq
\label{gamma}
\Gamma_{ij}(x)=f_{ij}+k\int A'^a_i A^a_j + k\int F^a_{ij}A^a_4
\qe

It follows from (\ref{ccr}) that for $\omega=exp\{i\omega^a(x,\sigma)T^a\}$
we  have  ($\partial_4=\partial/\partial\sigma$)

\eq
\label{a4}
A_4(x,\sigma)=-i\omega^\dag\partial_4\omega
\qe

Note ${\cal F}^\Omega$ is gauge invariant since a gauge transformation
$\Phi$ can 
be undone by a change of variables

\eq
\Omega(x) \rightarrow \Omega(x)\Phi^\dag (x)
\qe
 
Then (\ref{ccr}) implies from above (for two cocycle $\omega_2$) that
\cite{bkmco} 

\eqr
\alpha(x) \rightarrow \alpha(x)-\beta(x) +\omega_2(\omega,\phi)\\
\label{gtw}
\omega(x,\sigma) \rightarrow \omega(x,\sigma)\phi^\dag (x,\sigma)
\rqe

Hence from (\ref{gtw}) we have under gauge transformations

\eq
\label{gta3}
A_4(x,\sigma)\rightarrow\phi A_4 \phi^\dag +i\phi\partial_4\phi^\dag
\qe

We see that the only way that $\Omega$ couples to the other fields is
through the combination given by $A_4$.  In fact we see from the gauge
transformation given above that
$A_\mu=(A_i,A_4)$ forms a 4-dimensional vector field $\in M^3\times S^1$. 

\sect {Regularization; Wilson Tube}

One would naively expect that the trace of Wilson loops should constitute
all the gauge invariant quantities.  This does not hold for the Kac-Moody gauge
group since the trace gives a divergent result and is due to the fact that the
trace of any product of $Q^a(\sigma)'$s is infinity. It is essential that the
trace be regulated; to do so introduce the operator $L$ which is a
generalization of the Virasoro operator and is defined by

\eq
\label{reg}
L=\int Q^a(\sigma)f(\sigma-\sigma')Q^a(\sigma')
\qe

where $f>0$ and choose normalization $tr e^{-\beta L}=1$. The partition
function $tre^{-\beta L}$ yields a two-dimensional chiral version of the
WZW-model and has been studied in \cite{bkmc}. 

For $\beta>0$ we have 

\eq
\label{qqq}
tre^{-\beta L}Q^{a_1}(\sigma_1)Q^{a_2}(\sigma_2) .... Q^{a_n}(\sigma_n)<
\infty
\qe

The trace is performed over an irrep of the Kac-Moody algebra. The details
of the regulator L are unimportant in the continuum theory;
however this is not so for lattice Kac-Moody gauge fields. For the purpose
of deriving the continuum lagrangian we will only need that

\eq
\label{q} 
tre^{-\beta L}Q^a(\sigma) = 0
\qe

and

\eq
\label{qq}
\lim_
{\beta\rightarrow0}tre^{-\beta
L}Q^a(\sigma)Q^b(\sigma')={1\over\beta}\delta^{ab} H_{\sigma-\sigma'}
\qe

The regulator does not (cannot) commute with gauge transformations.  Hence
define the gauge invariant regulated {\it Wilson tube} starting at $x$ and
ending at $y$ $\in M^3$ by

\eq
\label{w}
W= tre^{-\beta L}\Omega(x)Pe^{i\int_{x}^{y}dz^i{\cal A}_i}\Omega^\dag(y)
\qe

There is another class of gauge invariant operators given by

\eq
\label{w1} 
V_{i..j}= tre^{-\beta L}\Omega(x)Pe^{i\int_{x}^{y}dz^i{\cal A}_i}{\cal
D}_i..{\cal D}_j\Omega^\dag(y)
\qe

where

\eq
\label{der}
{\cal D}_i=-i\partial_i+{\cal A}_i(y)
\qe

The operator $V$ is given by insertions at $y$.  For $x=y$ we have a marked
torus on which both $W$ and $V$ are defined.

We analyze the $\hat{U}(1)$ case.  In terms of
Fourier modes of $Q(\sigma)$ let  

\eq
\label{lu1}
L=\sum_{n=1}^{\infty}f_n Q_{-n}Q_n
\qe

To perform the trace we use the irrep of $\hat{U}(1)$ with $Q_0
|vac>=0$. The correlator for arbitrary $\beta$ is given by \cite{bkmc} 

\eqr
tre^{-\beta
L}Q_{-n}Q_n &=&k|n|{e^{-\beta |n| f_n}\over {1-e^{-\beta |n|f_n}}}\\
                &\sim& \lim_{\beta\rightarrow 0}{k\over{\beta f_n}}
\rqe

Hence we have from (\ref{qq}) and above

\eq
f_n={k\over H_n}
\qe

Consider a circle lying in the
$(1,2)$- plane with radius ${\bar R}$ and marked at $x_0$. $W$ is defined
on the marked {\it Wilson torus} and is given by 

\eqr
\label{wu1}
W &=& tre^{-\beta L}\Omega(x_0)Pe^{i\oint dz^i{\cal A}_i}\Omega^\dag(x_0)\\
  &=& e^{i\Delta} tre^{-\beta L}e^{i\int M(\sigma)Q(\sigma)}\\
  &=& e^{i\Delta}e^F
\rqe

where the magnetic flux at point $\sigma$ is given by

\eq
\label{delta}
M(\sigma)=\oint dz^i A_i(z,\sigma)
\qe

The phase $\Delta$ depends on $x_0$, and in cylindrical coordinates
$A_{i}=(A_{r},A_{\theta},A_{z})$ is given by 

\eqr
\Delta &=& \oint dz^i B_i + k\int M(\sigma)\omega'(x_0,\sigma)\nonumber\\
       &+& ik({{\bar R}\over{2\pi}})^2
\int_{0}^{R}d\sigma\int_{0}^{2\pi}d\nu  
A_{\theta}(\nu,\sigma)\int_{\nu}^{2\pi}d{\bar \nu}
A'_{\theta}({\bar \nu},\sigma)    
\rqe

and \cite{bkmc} 

\eqr
\label{f}
F&=&-{k\over 2}\sum_{n=1}^{\infty}n tanh({\beta n f_n\over 2})|M_n|^2 \\
 &\sim& \lim_{\beta\rightarrow 0}-{k\over 2 \beta}\sum_{n}{1\over f_n}|M_n|^2 
\rqe

Note the integration over the gauge field still has to be performed. Unlike
the usual Abelian field for which F is linear in the gauge field, for the
$\hat{U}(1)$ case the dependence is quadratic; this probably means that W
may not be the right order parameter for indicating the onset of confinement. 

\sect {Yang-Mills Lagrangian}

As a warm-up for the supersymmetric case we derive the lagrangian for the
pure gauge field. Define the lagrangian by

\eqr
\label{lkm}
{\cal L}_{KM}&=&-{1\over{4e^2}}\sum_{ij} tr e^{-\beta L}\Omega(x){\cal
F}^2_{ij}(x)\Omega^\dag(x)+{\cal L}'   \\
 &=&-{1\over{4e^2}}\sum_{ij} tr e^{-\beta L}(\Gamma_{ij}+\int
F^a_{ij}\rho^T_{ab}Q_b)^2+{\cal L}' \\
 &=&-{1\over{4e^2}}\sum_{ij}\{\Gamma_{ij}^2+{1\over\beta}\int
F^a_{ij}\rho^T_{ac}(x,\sigma)H_{\sigma -
\sigma'}F^b_{ij}\rho^T_{bc}(x,\sigma')\} +{\cal L}'
\rqe

where equation (\ref{qq}) has been used to perform the trace. The piece
${\cal L}'$ of the lagrangian is required for giving a kinetic term to $A_i$
in the $\sigma$ direction as well as for the dynamics of the $\Omega$ field .

From eqn (\ref{a4}) we have

\eq
\omega (x,\sigma)=Pe^{i \int_{0}^{\sigma} A^a_4 T^a}
\qe

and hence for $A^{\alpha \beta}_4=A_4^aC^{\alpha a \beta}$ we have

\eq
\label{rho}
\rho(x,\sigma)=Pe^{-\int_{0}^{\sigma}A_4}
\qe

Defining $g^2=e^2 \beta$ we have

\eq
{\cal L}_{KM}=-{1\over{4e^2}}\sum_{ij}\Gamma_{ij}^2-{1\over{4g^2}}\sum_{ij}\int
F_{ij}(x,\sigma)Pe^{-\int_{\sigma}^{\sigma'}A_4}H_{\sigma -
\sigma'}F_{ij}(x,\sigma') +{\cal L}'
\qe

We now derive ${\cal L}'$.  We have the natural gauge invariant interaction
(the 
coefficient is fixed by the requirement that one recover full 3+1 symmetry
for $k=0$)

\eq
{\cal L}'=-{1\over{2
g^2}}\sum_i\int F_{i4}(x,\sigma)Pe^{-\int_{\sigma}^{\sigma'}A_4}H_{\sigma-\sigma'}F_{i4}(x,\sigma') 
\qe

The lagrangian ${\cal L}_{KM}$ was derived in \cite{b1pl} using lattice theory 
arguments and without the explicit use of a regulator. The lagrangian (even
for the $\hat{U}(1)$ case) is asymmetric, nonlocal and nonlinear. The
function $H_\sigma$ at this point is arbitrary; we will fix it later by
demanding renormalizability.  

Note if we consider the special case of $H_{\sigma}=\delta(\sigma)$ (which
is one-loop renormalizable \cite{r}) we have 

\eqr
{\cal
L}_{KM}(x)&=&-{1\over 4e^2}\Gamma^2_{ij}(x)-{1\over4g^2}\int d\sigma
F^2_{\mu\nu}(x,\sigma)\\ 
         &=&{\cal L}_{central}+{\cal L}_{YM}      
\rqe

Note all the derivations so far hold in arbitrary dimensions; it is only when
supersymmetry is required that the choice of dimensions becomes
restricted. The theory is invariant under the $d=3$ Lorentz group $SL(2,R)$; for $k=0$ 
the bulk fields yield a Yang-Mills theory with the full $SL(2,C)$ symmetry of
$D=4$.  This pattern of enhanced symmetry will also be seen to hold for the
supersymmetric case.
 
We now rewrite  ${\cal L}'$ in a 3-dimensional form which has a
supersymmetric generalization.  For this we need the (suitably regularized)
Virasoro operator 

\eq
L_0=\frac{1}{C_{Adj}+2k}\int :Q^2(\sigma):
\qe

with

\eq
[L_0,Q_a(\sigma)]=i Q'_a(\sigma)
\qe

We then have the manifestly gauge invariant expression

\eq
\label{lprime}
{\cal L}'(x)={1\over{2g^2}}\sum_i tre^{-\beta L} [{\cal A}^\Omega_i(x),L_0]^2
\qe

Note ${\cal L}'$ is decoupled from the central extension due to the
commutator.  We hence have

\eq
{\cal L}_{KM}=-{1\over e^2}tre^{-\beta L}\{\Omega {\cal F}^2_{ij}\Omega^\dag-2[{\cal
A}^\Omega_i,L_0]^2\}
\qe

We use the background field method to quantize the theory.  Consider the
break-up 

\eq
{\cal A}_i={\cal C}_i +  a_i
\qe

where ${\cal C}_i$ is a classical field and $a_i$ the quantum
   field. We use a generalization of the 't Hooft gauge given by

\eq
\label{thooft}
\partial_ia_i +i[{\cal C}_i,a_i]=0
\qe

This gauge is invariant under background Kac-Moody gauge transformations on
${\cal C}_i$.  In components, for $a_i=b_i+\int a^\alpha_iQ^\alpha$ we have
from the equation above \cite{r}

\eqr
q^\alpha(x,\sigma)\equiv\partial_ia^\alpha_i - C_{\alpha \beta
\gamma}C^{\beta}_ia^\gamma_i=0 \\
    t(x)\equiv\partial_ib_i +k\int C'^\alpha_ia^\alpha_i = 0
\rqe

The quantum field theory is given by

\eq
Z=\int DaDb
e^{S[C+a]+S_{FP}}\prod_{x,\sigma,\alpha}\delta(q^\alpha(x,\sigma))\delta
(t(x))
\qe

where $S_{FP}$ is the Faddeev-Popov ghosts.  

It has been shown in \cite{r} that for $\hat{U}(1)$ $S_{FP}=0$. A one-loop
calculation for the 
$\hat{U}(1)$ case \cite{r} shows that the theory is renormalizable in D=3+1
if we take 

\eq
H_\sigma=\sum_{n=-\infty}^{\infty}e^{in\sigma}(1+a|n|)
\qe 

The coupling constant in the theory is
$\lambda={kg\over{e}}$ and which has a mass dimension of $(d-2)/2$; to make
this dimensionless in 3+1 we define $\bar{\lambda}={\lambda\over \surd R}$,
where $R$ is the radius of $S^1$.  The one loop beta function for
$\hat{U}(1)$ is given by \cite{r}

\eq
\beta=-{1\over{4a^2}}\bar{\lambda}^3
\qe

We see that the $\hat{U}(1)$ theory is asymptotically free! 

The limit of
$a\rightarrow 0$ is not uniform.  For $a=0$ we have
$H_{\sigma}=\delta(\sigma)$ ; the
theory is still renormalizable and $\beta=0$. 

If one takes limit ${R \rightarrow \infty}$, we see that (for $d>2$) $\bar{\lambda}\rightarrow0$; in effect this sets $k=0$ and the Kac-Moody
algebra reduces to the loop group algebra. 
  
\sect{Kac-Moody Chern-Simons Lagrangian}

We consider the usual Chern-Simons lagrangian to be the dimensional
reduction of the theory defined on $M^3\times S^1$ to $M^3$.  Define the
generalization of the Chern-Simons lagrangian with Kac-Moody gauge symmetry
by

\eq
\label{cs}
{\cal L}_{CS}=i\lambda\epsilon^{ijk}tre^{-\beta L}({\cal
F}^\Omega_{ij}{\cal A}^\Omega_k -i{2\over3}{\cal A}^\Omega_i{\cal
A}^\Omega_j{\cal A}^\Omega_k)
\qe

If one takes the bulk fields to be $\sigma$-independent ${\cal L}$
reduces to a $U(N)$ Chern-Simons theory in $d=3$.

For the $U(N)$ case it is known that for the theory  to be gauge invariant,
the coupling constant $\lambda$ is an integer (upto a constant).  For the
Kac-Moody case there is no such restriction on $\lambda$ since gauge
invariance is obtained explicitly by a coupling to the regulator field
$\Omega$; however, if one requires that the dimensionally reduced theory be
the usual Chern-Simons case $\lambda$ has to be similarly restricted. 

We work out the case of $\hat{U}(1)$ using a simpler lagrangian consisting
of only the first piece of ${\cal L}_{CS}$ given in (\ref{cs}) since
this is gauge invariant and reduces to the usual case. We then have

\eq
\label{cs1}
{\cal L}_{CS}=i\lambda\epsilon^{ijk} tre^{-\beta L}\Omega {\cal
F}_{ij}({\cal A}_k +i\partial_k)\Omega^\dag +{\cal L}'
\qe
 
Note that  unlike the case of (supersymmetric) Yang-Mills, the central
extension (phase) of the $\Omega$-field couples to the theory due to the
derivative acting on it; we  need
to have a kinetic term in the $\sigma$ direction and ${\cal L}'$ given in
eqn (\ref{lprime}) is added.  Working out the trace we obtain, for
$\Omega=e^{i\alpha(x)+\int\omega (x,\sigma)Q(\sigma)}$ the following

\eqr
{\cal L}_{CS}(x)&=&i\lambda \epsilon^{ijk}\{\int
F_{ij}(x,\sigma)H_{\sigma-\sigma'}A_k(x,\sigma') \nonumber \\
&+& \Gamma_{ij}(x)(B_k-\partial_k\alpha +k\int\omega'A_k-{k\over
2}\int\omega'\partial_k\omega)(x)\}+{\cal L}'(x)
\rqe 

with the kinetic term given by
\eq
{\cal L}'(x)=-{1\over{2 g^2}}\int
F_{i4}(x,\sigma)H_{\sigma-\sigma'}F_{i4}(x,\sigma')  
\qe

Since the kinetic piece ${\cal L}'$ doesn't couple
to the central extension there is no kinetic term for the $\alpha$ variable
and it appears as a Lagrange multiplier; integration over
it yields the constraint

\eq
\epsilon^{ijk}\partial_i\Gamma_{jk}=0
\qe

If we had used the lagrangian given by (\ref{cs}) we would have obtained

\eq
\label{cs2}
{\cal L}_{CS}\rightarrow{\cal L}_{CS}+{i\over
3}\lambda\epsilon^{ijk}(f^\Omega_{ij}-\Gamma_{ij})B^\Omega_k
\qe

Both $\hat{U}(1)$ lagrangians given in equations (\ref{cs1}) and
(\ref{cs2})  yield the same Chern-Simons 
lagrangian on dimensional reduction. Note that the lagrangian used
for the $\hat{U}(1)$ case in (\ref{cs1}) can also be used for the
non-Abelian case as it 
is fully gauge-invariant; of course this theory may not have much to do
with non-Abelian Chern-Simons as we have an extra $\Omega$-field in order to achieve
gauge invariance.  

We have obtained a generalization of the $U(1)$ Chern-Simons lagrangian to
$D=3+1$-dimensions with an extra parameter k. We need to check whether this
lagrangian is renormalizable, and whether the new nonlinearities yield any
new physics such as new classical solutions. A supersymmetric
generalization of ${\cal L}_{CS}$ can also be made based on the results of
the next section.

\sect{Supersymmetric Kac-Moody Gauge Fields }           

We obtain the action for the supersymmetric Kac-Moody gauge fields in
d-dimensions.  Note that it is necessary that supersymmetric Yang-Mills
exists 
in {\it both} $d$ and $D=d+1$ dimensions for the theory to be defined
without the introduction of extra fields.  The reason being, as we have seen
above, that the theory consists of a bulk field $A_\mu$ on a $D$-
dimensional manifold $M_d\times S^1$ and a boundary field $B_i$ in $d$
dimensions. The supersymmetry of the lagrangian will be limited to the
supersymmetry of the $d$ dimensional theory but we expect in the limit of
$k=0$ that we will have (enhanced) supersymmetry $D=4,N=1$ for the bulk
fields .   

Supersymmetric Yang-Mills exists only in  2,3,4,6 and 10
dimensions\cite{susyn},\cite{susyn1}; hence we can at most expect to have
supersymmetric Kac-Moody gauge fields in $d=2$ and $3$ dimensions.  

We start in $d=3$ with two
supercharges $Q_\alpha$, that is with $N=1,d=3$ \cite{susy3},\cite{susy3a},
\cite{susy3b}. Superspace
in $d=3$ is the extension of space $x$ to  superspace consisting of $(x,\theta)$ where
$\theta_\alpha$ is an anticommuting two component Majorana spinor.
Relevant formulae of supersymmetry in d=3 and 4 are briefly reviewed in the
Appendix. We choose pure 
imaginary Majorana $\gamma^i$ matrices with $\{\gamma^i,\gamma^j\}=\eta^{ij}$
where the metric has signature $(+,-,-)$; $\bar{\theta}=\theta^T\gamma^0$.

The $d=3$ real scalar multiplet superfield is dimensionless and  unconstrained,
and is given by (all superfields will be denoted by boldface).

\eq
{\bf s}(x,\theta)= A(x)+\bar{\theta}\psi(x)+{1\over 2}\bar{\theta}\theta
F(x) 
\qe

Note A is a real scalar field, $\psi$ is a Majorana spinor and F a real
auxiliary field.  For the Kac-Moody case introduce an (infinite) collection
of $d=3$ real scalar superfields ${\bf s}^a(x,\sigma,\theta)$ labeled by
$\sigma$ and define

\eq
{\bf S}(x,\theta)={\bf s} (x,\theta) +\int {\bf s}^a (x,\sigma,\theta)Q_a(\sigma)
\qe   

The Kac-Moody vector gauge field is described by a real spinor superfield
${\bf V}_\alpha$; in the Wess-Zumino gauge 

\eq
{\bf V}_\alpha(x,\theta)=i{\cal A}_i(\gamma^i\theta)_\alpha +{1\over
2}\bar{\theta}\theta \Psi_\alpha
\qe

where Kac-Moody fermions are given by

\eq
\Psi_\alpha(x)=\chi_\alpha(x)+\int \psi^a_\alpha(x,\sigma)Q_a(\sigma)
\qe

The gauge covariant field strength is given, in the Wess-Zumino gauge, by
the real spinor superfield

\eq
{\bf W}_\alpha(x,\theta)=\Psi_\alpha(x)+i{\cal
F}_{i}(\gamma^i\theta)_\alpha -{i\over2}{\bar \theta}\theta {\cal
D}_i\gamma^i\Psi_\alpha
\qe 

where ${\cal F}_i={1\over2}\epsilon^{ijk}{\cal F}_{jk}$ and

\eq
{\cal D}_i\Psi=\partial_i\Psi+i[{\cal A}_i,\Psi]
\qe

Note that unlike the case in $D=4$ both ${\bf V}_\alpha$ and ${\bf
W}_\alpha$ contain only physical degrees of freedom.  Supergauge
transformations in d=3 are given by $e^{i{\bf s}^a(x,\theta)T^a}$ and 
Kac-Moody supergauge transformations are given by  $e^{i{\bf S}(x,\theta)}$.

Supersymmetry transformations in $d=3$ are specified by a (real) Majorana
spinor $\epsilon$ and given by

\eqr
\label{1}
\delta {\cal A}_i&=&i\bar{\epsilon}\gamma_i\Psi\\
\label{2}
\delta \Psi &=&i{\cal F}_i\gamma^i\epsilon
\rqe

The bulk fields $\psi, A_i$ transform as components of a 3-dimensional
field for  each $\sigma$ whereas the the $U(1)$ boundary field $\chi$
transforms nontrivially (see Appendix).  
   
Let the regulator superfield be defined by

\eq
{\bf \Omega}(x,\theta)=e^{i{\bf \alpha}(x,\theta)+i\int{\bf
\omega}^a(x,\sigma,\theta)Q_a(\sigma)}
\qe

The regulator superfield is equivalent to local supergauge transformations in
$D=3+1$, and is given by  Lie group-valued local superfield as

\eq
\label{g}
{\bf g}(x,\sigma,\theta)=e^{i{\bf \omega}^a(x,\sigma,\theta)T^a}
\qe

Note that both ${\bf \Omega}$ and ${\bf g}$ are well defined as elements of
the Kac-Moody and compact Lie group respectively since in $d=3$ the scalar
superfield is real; in $D=4$ supergauge transformations in full generality
require the complexification of the group.
   
Let Lie algebra-valued superfield  be defined by

\eqr
{\bf
s}(x,\sigma,\theta)&=&-i{\bf g}^{\dag}{\partial\over{\partial\sigma}}{\bf g} \\
  &=&A_4(x,\sigma)+{\bar \theta}\xi(x,\sigma)+{1\over
2}\bar{\theta}\theta D(x,\sigma)
\rqe 

Superfield ${\bf s}^a(x,\sigma,\theta)$ can be considered to be an infinite
collection of $d=3$ real superfields or equivalently to be a single $D=4$ real
chiral superfield.  

Note under ordinary (not super) gauge transformations in $D=4$, we have
 ${\bf g}\rightarrow  {\bf g}\phi^\dag (x,\sigma)$ and which in turn yields

\eqr
A_4&\rightarrow& \phi A_4\phi^\dag +i\phi\partial_4\phi^\dag \\
\xi&\rightarrow& \phi\xi\phi^\dag \\
D &\rightarrow& \phi D\phi^\dag
\rqe

We see that superfield ${\bf s}(x,\sigma,\theta)$ has a bosonic component which
transforms like the $A_4$ component of a $D=3+1$ gauge field  and the 
fermionic and scalar fields transform like matter fields. Hence we can in
principle have gauge invariant couplings of these fields with the other
bulk fields in the system.  

Since ${\bf s}$ is a real scalar superfield under $d=3$ supersymmetry
transformations  we have \cite{susy3a},\cite{susy3b}

\eqr
\label{scalar1}
\delta A_4 &=& i\bar{\epsilon}\xi \\
\delta D &=& \bar{\epsilon}\gamma^i\xi \\
\label{scalar3}
\delta \xi_\alpha &=& i\partial_i A_4(\gamma^i\epsilon)_{\alpha} + D\epsilon_{\alpha}
\rqe   

We also have from eqns (\ref{1}) and (\ref{2})

\eqr
\label{vec1}
\delta A_i&=&i\bar{\epsilon}\gamma_i\psi\\
\label{vec2}
\delta \psi_\alpha &=&i F_i(\gamma^i\epsilon)_\alpha
\rqe

We check that fields $(A_\mu,\psi,\xi,D)$ with $d=3,N=1$  supersymmetry
transformation properties given above can be combined to yield the
supersymmetry transformation of $D=4, N=({1\over2},0)$. It can be shown
that for four component Majorana spinor given by 

\eq
\lambda ={1\over2}\pmatrix{\psi+i\xi \cr
                  \psi -i\xi} 
\qe

and with $N=({1\over2},0)$ supersymmetry (real) parameter given by

\eq
\alpha =\pmatrix{\epsilon \cr
                  \epsilon} 
\qe

the transformations (\ref{scalar1}) to (\ref{vec2}) for the $d=3$ case combine
to yield the  $D=4,N=({1\over2},0)$ supersymmetry 
transformations. As is expected, in the WZ-gauge all the derivatives are
covariantized \cite{susy4}, and the supersymmetry transformation
\cite{susy4},\cite{susy4a} is given by (see Appendix) 

\eqr
\delta A_\nu &=& \bar{\alpha}\Gamma_\nu\lambda \\
\delta D &=& \bar{\alpha}\Gamma_5\Gamma_\nu{\cal D}^\nu\lambda \\
\delta\bar{\lambda} &=&
-\bar{\alpha}\Sigma_{\mu\nu}F^{\mu\nu}+i\bar{\alpha}\Gamma_5 D
\rqe   

We see from above that the $D=4, N=({1\over2},0)$ superalgebra is a tensor
product of 
the expected  $d=3,N=1$ subalgebra times the subalgebra consisting of
the supersymmetry transformation of the $d=3,N=1$ scalar superfield
$(A_4,\xi,D)$.

The full $D=4, N=1=({1\over2},{1\over2})$ superalgebra with four supercharges
\cite{susy4b} has an arbitrary Majorana 
fermion for $\alpha$ given by a two-component complex spinor $\zeta_\alpha$
such that 

\eq
\alpha =\pmatrix{\zeta \cr
                  \zeta^\ast} 
\qe

Similar to the pure bosonic case, we expect that for the supersymmetric
case the bulk fields will yield a $D=4, N=1$ supersymmetric Yang-Mills
lagrangian and the boundary fields will couple to the bulk fields, and will
yield a  $d=3,N=1$ supersymmetric lagrangian. 
The boundary fields have the required degrees of freedom; the 
bulk fields $(\psi,A_i)$ together  with the regulator (bulk) fields
$(\xi,A_4,D)$ 
have the exact field content required to make up a $D=3+1, N=1$ hermitian
vector multiplet.

In analogy with the bosonic case we define the $d=3$ supersymmetric
lagrangian by 

\eq
\label{susykm}
{\cal L}_{SKM}(x)=-{1\over e^2}\int d^2\theta tre^{-\beta L}\{{\bf \Omega}{\bf \bar{W}}{\bf
W}{\bf \Omega}^{\dag} - 2[{\bf \bar{V}}^{\bf
\Omega},L_0][{\bf V}^{\bf \Omega},L_0]\}
\qe

Note by construction the lagrangian is supergauge invariant.   

To compute ${\cal L}_{SKM}$ note from the Kac-Moody algebra we have

\eq
{\bf \Omega}{\bf W}_\alpha{\bf \Omega}^\dag={\bf
U}_\alpha(x,\theta)+\int{\bf W}^a_\alpha\rho_{ab}(\omega)Q_b
\qe

Note ${\bf U}_\alpha(x,\theta)$ is the supersymmetric generalization of the
bosonic 
$\Gamma_{ij}$-term and is given by 

\eq
{\bf U}_\alpha(x,\theta)={\bf w}_\alpha(x,\theta)+k{\bf C}_\alpha(x,\theta)
\qe

where the central terms are given by

\eq
{\bf C}_\alpha =-{i\over 2}\epsilon^{ijk}\int
A'^a_jA^a_k(\gamma^i\theta)_\alpha
-{1\over2}\bar{\theta}\theta\int A'^a_i(\gamma^i\psi^a)_\alpha +\int{\bf
W}^a_\alpha{\bf s}^a
\qe

and the $d=3$ spinor  superfield is  given by
   
\eq
{\bf
w}_\alpha
=\chi_\alpha(x)+i\epsilon_{ijk}(\gamma^i\theta)_\alpha\partial^jB^k(x)
-{i\over2}{\bar \theta}\theta \partial_i\gamma^i\chi_\alpha(x)
\qe 

To simplify the expression for the lagrangian we use a regularization which
is local, that is

\eq
\lim_
{\beta\rightarrow0}tre^{-\beta
L}Q^a(\sigma)Q^b(\sigma')={1\over\beta}\delta^{ab} \delta(\sigma-\sigma')
\qe

We hence obtain the action

\eqr
{\cal S}&=& -{1\over e^2}\int d^3xd^2\theta \bar{\bf U}(x,\theta){\bf
U}(x,\theta) \nonumber \\
&-& {1\over 4 g^2} \int d^3xd\sigma
 \{F^2_{\mu\nu}+{i\over2}\bar{\lambda}\Gamma^\mu{\cal D}_\mu\lambda
       +{1\over2}D^2\}    \\ 
    &=& {\cal S}_{central} +{\cal S}_{SYM}
\rqe

The full supersymmetric theory consists of a $D=4,N=1$ supersymmetric
Yang-Mills theory given by ${\cal S}_{SYM}$ coupled to a $U(1)$
boundary field with $d=3,N=1$ supersymmetry. The complete action ${\cal
S}$  has $D=4,N=({1\over2},0)$ supersymmetry.  

For the  case of $k=0$ we obtain a higher $D=4,N=1$ supersymmetry for the
Yang-Mills bulk
fields whereas the (decoupled) boundary fields continue to have $d=3,N=1$
supersymmetry.   

A consequence of our derivation, in particular the use of the regulator
field $\Omega$, is that for the $D=4,N=1$ Super Yang-Mills fields we can choose
the {\it superaxial} gauge given by ${\bf s}^a(x,\sigma,\theta)=0$, or in
components 

\eq
A_4=\xi_{\alpha}=D=0
\qe 

\sect {Conclusions}

We have generalized of the  group of gauge symmetry to that of infinite
dimensional groups. The key feature of regularizing the trace yields many new
features in the lagrangian including new nonlinear, nonlocal and asymmetric
interactions.  Kac-Moody (super)gauge transformations act nontrivially on
the $d$ 
dimensional boundary fields, whereas on the $D=d+1$ bulk fields these are
simply the usual $D$ dimensional local (super)gauge transformations.  

The cases of Yang-Mills, Chern-Simons and supersymmetric gauge fields all
have new features.

The supersymmetric generalization for the loop group case of $k=0$
is interesting in its own right since we can set the radius of the extra
dimension to infinity.  This then provides 
a way of obtaining a $D=4,N=1$ supersymmetric Yang-Mills theory
starting from a $d=3$ dimensional supersymmetric theory.   

In the Kaluza-Klein
approach one starts in $D=d+1$ dimensions and reduces the
theory to $d$ dimensions by compactification.  The loop group approach is
the inverse of the Kaluza-Klein approach since   we
started with a $d$ dimensional theory and 'lifted' it  to $D=d+1$. It was
assumed that the system has the infinite  
dimensional loop group as its gauge symmetry; the continuous space $S^1$
underlying the loop 
group was then interpreted as an (extra) space dimension; the extra degrees of
freedom in a (super) gauge transformation provided the extra
fields required for increasing the dimension  from $d$ to $D=d+1$. 

Loop group supergauge transformations in  
$d$ dimensions become the usual local supergauge transformations in
$D=d+1$. This  
programme was carried out in arbitrary dimensions for the bosonic case.
However for 
the supersymmetric case one increased  dimension from 3 to 3+1
(without adding  extra fields by hand) since
both of these dimensions allow for the existence of supersymmetric gauge
fields.      

\vskip 0.3cm

\begin{center}
{\bf Acknowledgments}
\end{center}

I thank Rajesh Parwani and Yim Kok Kean for helpful
discussions during the early part of this work, and   
Michael Spalinski, Ansar Fayyazuddin and Peter Cho for useful discussions.
I also would like to thank  
Cumrun Vafa and the string theory group for their kind hospitality.

\vskip 0.3cm
\noindent

\section*{Appendix}

\appendix

\sect{Supersymmetry in d=3}

\vskip 0.3cm

Superspace in $d=3$ is given by $(x,\theta)$, where $\theta_{\alpha}$ is a real
two-component Majoran spinor; we have pure imaginary Majorana
gamma-matrices given by 

\eq
\gamma^0\equiv\gamma^1=i \pmatrix{0&-1 \cr 1&0\cr} , \gamma^2=i \pmatrix{0 & 1
 \cr 1 &0 \cr} ,
                               \gamma^3=i \pmatrix{1 & 0 \cr 0 & -1 \cr}
\qe                    

The supercharge ${\bf Q}_\alpha$ and superderivative ${\bf D}_\alpha$ are defined by

\eqr 
{\bf Q}_\alpha&=&{\partial\over{\partial\theta^\alpha}} +
i(\gamma^i\theta)_\alpha \partial_i \\
{\bf D}_\alpha&=& {\partial\over{\partial\theta^\alpha}}
-i(\gamma^i\theta)_\alpha \partial_i  
\rqe
Supersymmetry transformations are given by real two-component spinor
$\epsilon_\alpha$; for a real scalar superfield we have 

\eq
\delta{\bf s}=\bar{\epsilon}{\bf {Qs}}
\qe

with $\bar{\epsilon}=\epsilon^T\gamma^0$. Eqns
(\ref{scalar1})-(\ref{scalar3}) are obtained from the transformation above.

Kac-Moody superfield transformations given in (\ref{1}) and (\ref{2}) yield
(\ref{vec1}) and (\ref{vec2}) for the bulk fields; for the boundary fields
this yields

\eqr
\delta \chi_\alpha&=&{i\over2}\epsilon^{ijk}(f_{ij}+k\int
A'_iA_j)(\gamma_k\epsilon)_\alpha \\
\delta B_i&=&\bar{\epsilon}\gamma_i\chi
\rqe

The pure imaginary Majorana representation of the $D=4$ gamma matrices is
given by \cite{susy4}

\eqr
\Gamma_i=\pmatrix{0 &\gamma_i \cr \gamma_i & 0 \cr},\Gamma_4=i \pmatrix {1 & 0
\cr 0 & -1 \cr} = i\Gamma_5 
\rqe

For $\Sigma_{\mu\nu}={1\over2}[\Gamma_\mu,\Gamma_\nu]$ we have

\eqr
\Sigma_{ij}&=&{1\over2} \pmatrix
{0&[\gamma_i,\gamma_j]\cr[\gamma_i,\gamma_j]&0\cr} \\
\Sigma_{i0}&=&\pmatrix{0&-\gamma_i\cr \gamma_i&0\cr}
\rqe

The charge conjugation matrix $C$ is given by

\eqr
C\Gamma_\mu C^{-1}=-\Gamma_\mu^T \\
C^T=-C
\rqe

For the Majorana realization we  have

\eq
C=\pmatrix{\gamma_0 & 0 \cr 0 &\gamma_0 \cr}  
\qe

Let $\psi_\alpha$ and $\xi_\alpha$ be two-component real spinors;  $D=4$
Majorana fermion $\lambda$ is given by
\eq
\lambda={1\over2}\pmatrix{\psi+i\xi \cr \psi - i\xi \cr}
\qe

We then have ($\Gamma_0\equiv\Gamma_1$)

\eqr
\bar{\lambda}&=&\lambda^\dag\Gamma_0=\lambda^TC\\
              &=&\pmatrix{\bar{\psi}+i\bar{\xi} & \bar{\psi}-i\bar{\xi} \cr}
\rqe

The spinor field-tensor ${\bf W}_\alpha$ is given by \cite{susy3}

\eq
{\bf W}_\alpha={1\over2}{\bf D}^\beta{\bf D}_\beta{\bf V}_\alpha +
{i\over2}[{\bf V}^\beta,{\bf D}_\beta{\bf V}_\alpha]+{1\over6}[{\bf
V}^\beta,\{{\bf V}_\beta,{\bf V}_\alpha\}]
\qe

Supegauge transformations are given by 

\eq
{\bf g}=e^{i{\bf s}^a(x,\theta)T^a}
\qe

and

\eqr
{\bf V}_\alpha&\rightarrow&{\bf g}({\bf V}_\alpha+i{\bf D}_\alpha){\bf
g}^\dag \\
{\bf W}_\alpha&\rightarrow& {\bf gW}_\alpha{\bf g}^\dag
\rqe

Let ${\bf W}_\alpha$ be given in an arbitrary gauge; then we have

\eq
{\bf W}_\alpha^{WZ}={\bf g}_{WZ}{\bf W}_\alpha{\bf g}^\dag_{WZ}
\qe

and similarly for ${\bf V}_\alpha$.  We see from above that we can always
work  in the Wess-Zumino gauge as long as we are computing gauge-invariant
quantities; in our case since the regulator field ensures manifest
gauge-invariance we can work in the WZ-gauge without any loss of generality.


\newpage
	
\begin{thebibliography}{99}

\bibitem{b1pl} B.E.Baaquie Phys Lett {\bf 271B} (1991) 343
\bibitem{b2pr} B.E.Baaquie Phys Rev {\bf D53} (1996) 879; hep-th/9511094
\bibitem{r}    B.E.Baaquie and R.Parwani Phy Rev {\bf D54} (1996) 5259
;hep-th/9511165
\bibitem{bkm}  B.E.Baaquie Phys Lett {\bf 177B} (1986) 310 
\bibitem{bkmco}B.E.Baaquie Mod Phys Lett A Vol6, No.31 (1991) 2881
\bibitem{bkmc} B.E.Baaquie Mod Phys Lett A Vol8, No.26 (1993) 2449
\bibitem{bu1km}B.E.Baaquie Mod Phys Lett A Vol7, No.13 (1992) 2973
\bibitem{susyn}M.Green, J.Schwarz and E.Witten ``Superstring Theory'' Cambridge
Univ Press (1984)  
\bibitem{susyn1}J.D.Lykken ``Introduction to Supersymmetry''; hep-th/9612114
\bibitem{susy3} S.J.Gates, M.T.Grisaru, M.Rocek and W.Siegel ``Superspace, or
One Thousand and One Lessons in Supersymmetry'' Benjamin Cummings (1983)
\bibitem{susy3a} I.Affleck, J.Harvey and E.Witten Nucl Phys {\bf B206} (1982)
413 
\bibitem{susy3b}D.Kabat and Soo-Jong Rey Nucl Phys {\bf B508}(1997) 535 ;
hep-th/9707099 
\bibitem{susy4} J.Wess and J.Bagger ``Supersymmetry and Supergravity''
Princeton University Press (1992) 
\bibitem{susy4a}P.West ``Introduction to Supersymmetry and
Supergravity'' World Scientific (1990)
\bibitem{susy4b}L.Alvarez-Gaume and S.F.Hassan
``Introduction to S-Duality in N=2 Supersymmetric Gauge Theories''
hep-th/9701069 
 
 
\end{thebibliography}
\end{document}